\documentclass[twocolumn,showpacs,prl,preprintnumbers,amsmath,amssymb]{revtex4}

\usepackage{graphicx}
\usepackage{dcolumn}
\usepackage{bm}

\begin{document}

\preprint{Physical Review Letters, 96, 024501, 2006}

\title{Predicting the progress of diffusively limited chemical reactions \\in the presence of chaotic advection}

\author{P.E. Arratia}
\author{J.P. Gollub}%
\affiliation{%
Department of Physics, Haverfod College, Haverford, PA
19041\\Department of Physics, University of Pennsylvania,
Philadelphia, PA 19104
}%

\date{\today}

\begin{abstract}
The effects of chaotic advection and diffusion on fast chemical
reactions in two-dimensional fluid flows are investigated using
experimentally measured stretching fields and fluorescent monitoring
of the local concentration. Flow symmetry, Reynolds number, and mean
path length affect the spatial distribution and time dependence of
the reaction product.  A single parameter
\emph{$\overline{\lambda}N$}, where \emph{$\overline{\lambda}$} is
the mean Lyapunov exponent and \emph{N} is the number of mixing
cycles, can be used to predict the time-dependent total product for
flows having different dynamical features.
\end{abstract}

\pacs{47.52.+j, 82.40.Ck, 47.70.Fw, 05.45.-a}
\maketitle

Chemical reactions in solution are enormously enhanced by mixing,
which brings the components into intimate contact along lines or
surfaces.  While turbulent flows are effective in producing mixing,
even laminar velocity fields can produce complex distributions of
material and promote reaction through "chaotic advection", in which
nearby fluid elements separate exponentially in time.  For fast
reactions, the overall rate is determined by diffusion of reactants
into the reaction zones, a process that is augmented by the
stretching of fluid elements.

The interplay of stretching, diffusion, and reaction has been
investigated in various one-dimensional
(1D)~\cite{ranz79,Muzzio89,Sokolov91} and 2D
models~\cite{Muzzio96,Toro98,Sokolov00,Giona02,kari04} in both open
and closed domains. For irreversible, fast chemical reactions,
numerical studies have attempted to relate product concentration
growth to the stretching properties of the flow.  At early times,
and assuming uniform stretching leading to exponentially growing
reactant interfaces, the product growth is also expected to be
exponential~\cite{Chella84,Sokolov91b,Chertkov03}. Several numerical
investigations have shown that inhomogeneous stretching should
strongly affect local and overall reaction progress
~\cite{Chella84,Muzzio96,Szalai03}.  At later times, and using the
passive scalar approximation in the presence of diffusion (valid for
an infinitely fast reaction), several different asymptotic
functional forms for the product growth have been
predicted,~\cite{Sokolov91,Tang96,Sokolov00,Giona02a}. Boundary
effects may lead to different regimes as time
evolves~\cite{Chertkov03}.

However, there has been little opportunity to test theoretical
models because of the difficulty of measuring the stretching
properties of experimental flows.  Here, we use a recently developed
method to measure stretching fields of two-dimensional time-periodic
flows~\cite{Voth02}, which quantitatively describe the local finite
time Lyapunov exponent field, and apply it to a fast reaction in
order to address a fundamental question:  How does chaotic mixing
influence the spatial distribution of the reaction product, and the
time dependence of product growth?  While locally the reaction
progress is related to the stretching properties of the flow, we
show that the spatial average of the product concentration, which is
proportional to the total quantity of product, depends on time and
on the mean Lyapunov exponent in a way that is independent of the
spatial structure of the flow (c.f. Fig. 4).  We give a simple
function describing this dependence.
\begin{figure}
\includegraphics[scale=0.9]{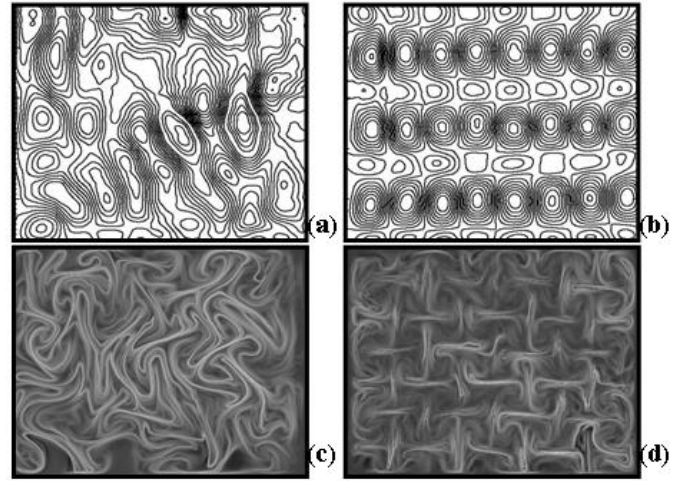}
\caption{\label{fig:epsart} Streamlines computed from measured
velocity fields at \emph{Re}=56 and path length \emph{p}=2.5 for (a)
disordered and (b) ordered magnet array configurations.  The ordered
case shows clear lines of mirror symmetry.  Corresponding stretching
fields for (c) disordered and (d) ordered flows, showing the
magnitude of stretching over an interval \emph{$\Delta$t}= 1 period.
Regions of strong stretching are localized in both cases.}
\end{figure}
The experimental configuration, described in more detail
elsewhere~\cite{Voth03}, is a thin fluid layer (1 mm thick)
containing the reactants, that overlies a somewhat deeper conducting
fluid layer (3 mm thick). Chaotic flows are created by
magneto-hydrodynamic forcing of the conducting fluid layer. A
time-periodic electric current (with frequency \emph{f} in the range
10-140 mHz) passing through the lower layer, in the presence of an
array of magnets, drives a time-dependent vortex flow that may be
either spatially ordered or disordered depending on the magnet
arrangement. The fluid is a 20\% glycerol-water mixture (viscosity =
1.74 cP, density   = 1.1 g/cm$^{3}$), about 10 cm x 10 cm in size,
but only the central 8 x 8 cm is imaged.  The flow in the upper
layer is nearly two-dimensional~\cite{Paret97} and typical RMS
velocities (\emph{U}) are 0.05 to 0.7 cm/s. The Reynolds number,
\emph{Re}= \emph{$\rho$LU}/$\mu$, based on the mean magnet spacing
(\emph{L}=2 cm) is in the range 5 to 75.  The path length parameter,
\emph{p}=\emph{U}/\emph{Lf}, which describes the mean displacement
of a typical fluid element in one forcing period, is in the range
0.5 to 3.5.

We investigate reactive mixing using an aqueous acid-base reaction
NaOH + HCl $\rightarrow$ NaCl + H$_{2}$O, or H$^{3}$O+ + OH$^{-}$
$\rightarrow$ 2H$_{2}$O, in the upper fluid layer.  We write it
schematically as A + B $\rightarrow$ 2P.  The reaction is fast and
second order.  Its rate constant is \emph{k}=1.1 x 10$^{8}$ M$^{-1}$
s$^{-1}$, and the reaction speed is characterized by the
Damk\"{o}ler number \emph{Da}= \emph{kC$_{0}$L$^{2}$/D}, based on
the ratio of the diffusion time scale (\emph{L$^{2}$}/\emph{D}) to
the reaction time scale (\emph{kC$_{0}$}). Here, the diffusion
constant \emph{D} of either acid or base with its counter-ion is
about 10$^{-5}$ cm$^{2}$/s, and \emph{C$_{0}$}=2.2 x10$^{-2}$ M is
the initial reactant concentration.  We note that \emph{Da} is large
($>$ 10$^{5}$), so that the reaction is limited by the diffusion
fluxes toward the lines of contact. However, diffusion is enhanced
by the stretching of interfaces.  The relative importance of
stretching and diffusion is given by the (Lagrangian) Peclet number
\emph{Pe}=\emph{L$^{2}$$\overline{\lambda}$/D}, where
\emph{$\overline{\lambda}$} is the mean Lyapunov exponent of the
flow. Here, \emph{Pe} is typically large and in the range 3.48 x
10$^{5}$ to 1.02 x10$^{7}$.

To determine fluid stretching, we first obtain high resolution
velocity fields using particle tracking methods described
elsewhere~\cite{Voth02}. Streamlines of the measured velocity fields
for both ordered and disordered magnet array configurations at
\emph{Re}=56 (and \emph{p}=2.5) are shown in Figs. 1(a,b),
respectively.  The disordered flow has no spatial symmetry while the
ordered flow has reflection and discrete translation symmetry along
the coordinate axes. Regular (non-mixing) regions are found in the
ordered flow at a given \emph{Re}, and in the disordered flow at low
\emph{Re}.

Next, we use the velocity fields to construct displacement maps over
a selected time interval \emph{$\Delta$t}, and then compute
stretching fields from the maps by differentiation [15].  This
method, originally developed in numerical studies, provides high
resolution stretching fields~\cite{Haller00,Haller01}.  Examples are
shown in Fig. 1(c,d). The local stretching (\emph{S}) measures the
deformation of an infinitesimal circular fluid element located
initially at (\emph{x,y}) over the interval \emph{$\Delta$t}.  The
local finite time Lyapunov exponent is defined as
\emph{$\overline{\lambda}$}=(log \emph{S})/\emph{$\Delta$t}.
Stretching fields computed over 1 period, corresponding to the
disordered and ordered array at \emph{Re}=56 (\emph{p}=2.5), are
shown in Figs. 1(c,d), respectively. Both fields show a wide
distribution of stretching values~\cite{Arratia05} stronger at some
locations than at others by a factor of 1000.  In the ordered case
(Fig 1d), it is clear that the stretching is highly inhomogeneous,
being much larger along lines passing through stagnation
(hyperbolic) points of the flow than in other regions. The strongest
overall stretching occurs for the disordered array (Fig. 1c), due in
part to the lack of spatial symmetry of the velocity field.  We also
find stronger stretching for flows with larger \emph{p} at a given
\emph{Re} and magnet array configuration.  The mean Lyapunov
exponent \emph{$\overline{\lambda}$} is computed by averaging the
local values over the image domain.

We now show in Fig. 2 the reaction of initially segregated aqueous
solutions containing reactant A (acid) and reactant B (base) at
\emph{Re}=37 and \emph{Re}=56 for the disordered flow, and also at
\emph{Re}=56 for the ordered flow.  Initially, a solid barrier (the
dotted line in Fig. 2a) separates the acid, which contains a pH
sensitive fluorescent dye, from the base.  The barrier is lifted and
the reaction is observed over varying numbers \emph{N} of cycles.
(\emph{N}=10 and \emph{N}=30 are shown in the upper and lower rows
of Fig. 2.)

We define normalized concentration fields for the acid \emph{A},
base \emph{B}, and product \emph{P}, as follows:
\emph{$\widetilde{A}$}=\emph{A}/\emph{A$_{0}$},
\emph{$\widetilde{B}$}=\emph{B}/\emph{B$_{0}$}, and
\emph{$\widetilde{P}$}=\emph{P}/\emph{P$_{final}$}, where
\emph{A$_{0}$} and \emph{B$_{0}$} are the initial reactant
concentrations, and \emph{P$_{final}$} is the product concentration
in the fully reacted state. Using separate calibration experiments,
we determine \emph{$\widetilde{A}$$(x,y,t)$} from the local light
intensity. From the conservation of material expressed in the
statement A + B $\rightarrow$ 2P, it can be shown that averaged over
the entire cell
$\langle\widetilde{P}\rangle$=1-2$\langle\widetilde{A}\rangle$,
where $\langle A\rangle$=1/\emph{n}$\sum_{j=1}^{n} A_{j}$, and
\emph{n} is the number of pixels in the image. Though this relation
is not accurate locally due to advection and diffusion, we use
\emph{$\widetilde{P}$}=1-2\emph{$\widetilde{A}$} as an approximation
to display the local normalized product concentration field.
\begin{figure}
\includegraphics[scale=1.0]{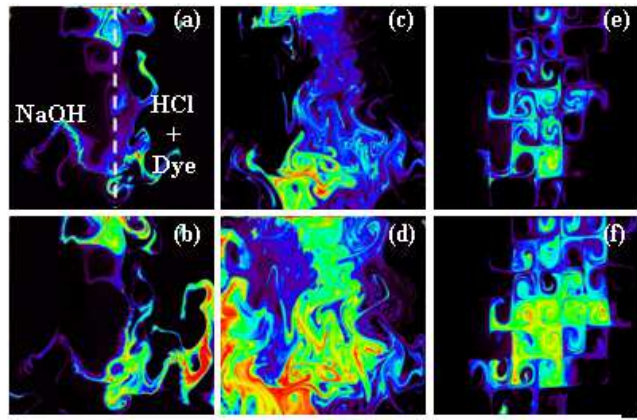}
\caption{\label{fig:epsart} (Color online) Normalized product
concentration fields ($\widetilde{P}$) vs. number of periods
(\emph{N}) at \emph{p}=2.5.  (a) \emph{Re}=37 (disordered),
\emph{N}=10; (b) Same as (a) for \emph{N}=30; (c) \emph{Re}=56
(disordered), \emph{N}=10; (d) Same as (c) for \emph{N}=30; (e)
\emph{Re}=56 (ordered), \emph{N}=10; (f) Same as (e) for
\emph{N}=30. Fully reacted regions appear in red, as shown in the
color bar. Unreacted regions are shown in black.}
\end{figure}
As time evolves, the interplay of stretching, diffusion, and
reaction creates a complex pattern, with regions of high (red) and
low (dark) normalized product concentration.  The development of
convoluted interfaces is evident early.  At longer times, systematic
differences between the different flows become clear.  For
\emph{Re}=37 (disordered array), the reacted regions are spatially
extended, but large unreacted regions remain even after 30 periods.
 This flow departs less from time-reversibility, a necessary
condition for mixing~\cite{Voth03}, due to its lower inertia than
does the \emph{Re}=56 case. Even in the reacted regions, the product
concentration is small at lower \emph{Re}.

Product formation is also affected by the extent of regular
(non-mixing) islands, which are favored by the mirror symmetry of
the ordered flow (Fig. 2e,f), but that also occur for the disordered
flow at the lower Re (Fig. 2a,b).  These are regions of low
stretching where the interface between reactants grows approximately
linearly in time, and the local reaction rate is less than in
regions of high stretching. Such isolated regions are not visible at
in Fig. 2(c,d) (higher \emph{Re}=56, disordered flow), where the
accumulated product is higher than it is in the ordered flow for the
same Re and reaction time. These qualitative differences show that
the product distribution is substantially affected by the flow
pattern.
\begin{figure}
\includegraphics[scale=0.95]{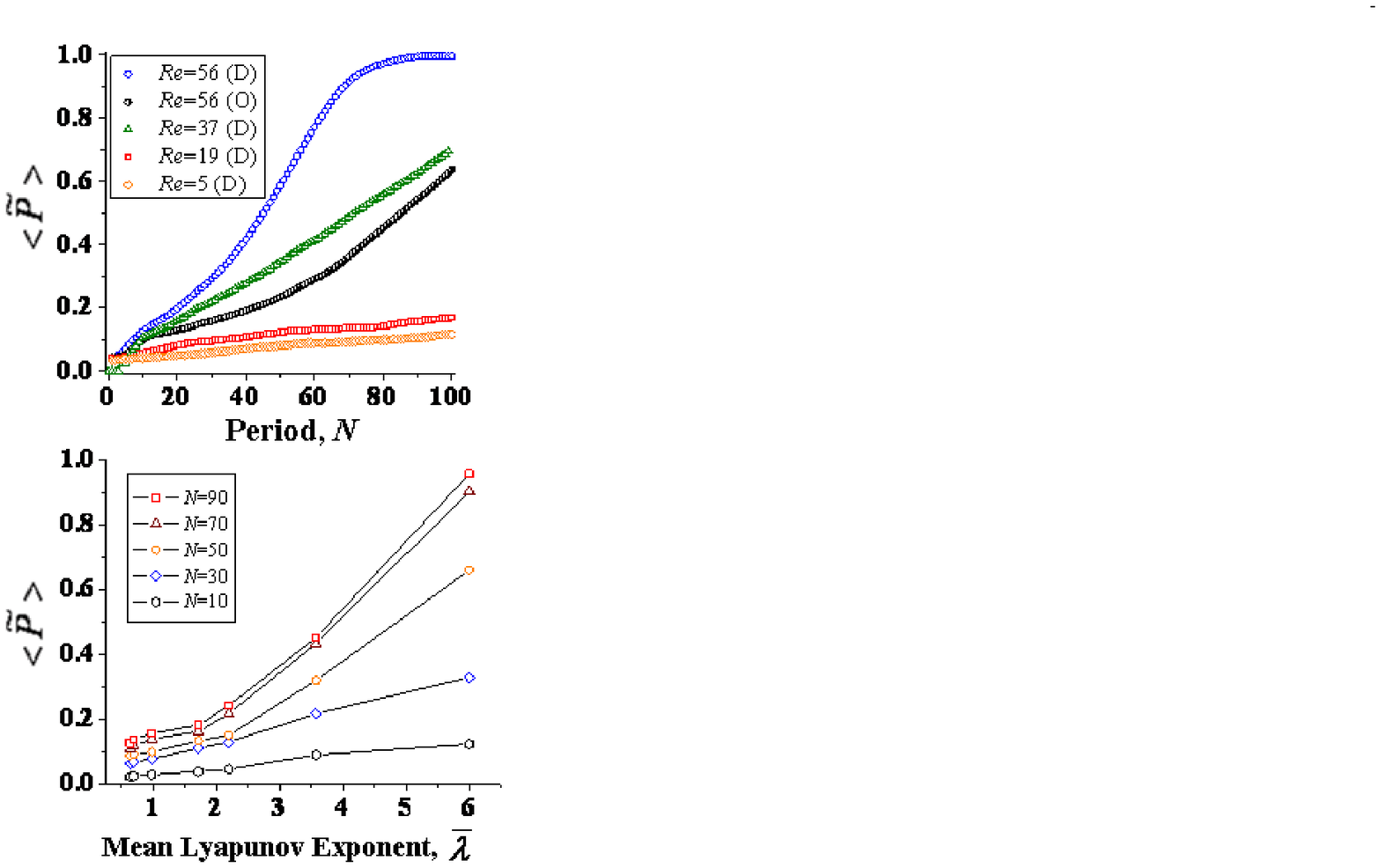}
\caption{\label{fig:epsart} (Color online) (a) Spatial averaged
normalized product concentration $\langle\widetilde{P}\rangle$ as a
function of time (or \emph{N}) and \emph{Re} for several different
flows (D=disordered, O=ordered, all at \emph{p}=2.5). (b)
$\langle\widetilde{P}\rangle$ vs. mean Lyapunov exponent (
\emph{$\overline{\lambda}$}) for various \emph{N} (disordered flows
for various \emph{Re}). Lines are added to guide the eye. Product
growth is slow for small  but accelerates for large   at later
times.}
\end{figure}

Next we explore the evolution of the spatial average of the
normalized product $\langle\widetilde{P}\rangle$ (or the total
quantity of product) as a function of time for flows having
different dynamical features. Fig. 3(a) shows that the high
\emph{Re} and \emph{p} disordered flow produces the fastest product
growth, while lower \emph{Re}, \emph{p} or spatial order
substantially reduces the growth rate. The reason is that
irreversibility due to inertia (large \emph{Re}), large particle
displacement (\emph{p}), and the absence of barriers to transport
(disorder) are required for rapid mixing. For all flows, the initial
product growth is roughly linear rather than exponential in time.
This is especially clear for low \emph{Re} and ordered flows, where
there are substantial low stretching regions. Although the simplest
theories suggest exponential early growth~\cite{Chertkov03} this is
not apparent in the data, most likely due to the wide distribution
of local stretching rates first pointed out in Ref.~\cite{Voth02}. A
numerical study in 2D flows~\cite{Muzzio96} also finds that
non-chaotic regions lead to sub-exponential initial product growth.

The average normalized product $\langle\widetilde{P}\rangle$ is most
usefully parameterized by the mean Lyapunov exponent
\emph{$\overline{\lambda}$} rather than \emph{Re} or \emph{p}. In
Fig. 3(b), we show the variation of $\langle\widetilde{P}\rangle$ as
a function of \emph{$\overline{\lambda}$} at different numbers of
periods \emph{N} for disordered flows with different \emph{Re}
(\emph{p}=2.5). The growth with \emph{$\overline{\lambda}$} is
approximately linear for small \emph{$\overline{\lambda}$} but
accelerates for large \emph{$\overline{\lambda}$} at the later
times.

Although the concentration patterns are complex, the evolution of
$\langle\widetilde{P}\rangle$ (at least after an initial transient)
turns out to be a function only of \emph{N} and
\emph{$\overline{\lambda}$}.  In Fig. 4, we plot
$\langle\widetilde{P}\rangle$ vs. \emph{$\overline{\lambda}N$} for
the disordered flows at various \emph{Re} and \emph{p}.  We find
that product concentration curves collapse onto a single master
curve with no adjustable parameters. This is a surprising result
because even though the flows have the same magnet array
configuration, they possess different degrees of time reversibility
and mean particle displacements per cycle, and regular regions occur
to a different extent in the various flows. In the insert to Fig. 4,
we show that this scaling behavior can be extended to ordered flows
as well. Remarkably, the rescaled ordered flow data for different
\emph{Re} fall onto the same master curve as the disordered array
cases, which include various \emph{Re} and \emph{p}.

The dependence of product concentration on and time has been
recently considered theoretically~\cite{kari05} using a simplified
model for the reaction interface.  The authors suggest that after a
short time (\emph{t}$<$1/\emph{$\overline{\lambda}$}) the normalized
concentration should be described by
\begin{eqnarray}
\label{eq1}
 \langle\widetilde{P}\rangle=1-exp({-a{\overline{\lambda}}N})
 \end{eqnarray}
where \emph{a} is a constant.  A similar relationship is also found
in a numerical study using periodic boundary conditions for 2D
flows~\cite{Giona02}.  This result, shown by the dotted line in Fig.
4 for the best fitting value of a describes our data adequately for
$\emph{$\overline{\lambda}N$}<300$.  However, for larger values,
$\langle\widetilde{P}\rangle$ grows more rapidly than this form
predicts.

Other theoretical works show a double exponential growth, or a more
complex spatio-temporal behavior depending on the reaction
rate~\cite{Sokolov91,Tang96} or the presence of
boundaries~\cite{Chertkov03}. Finally, one might consider a solution
of Floquet form consisting of a sum of several
exponentials~\cite{Haller04}, since the flow is periodically driven.
However, none of these forms is found to fit the data over the
entire range of \emph{$\overline{\lambda}N$}.

We can fit the entire data set (ordered, disordered, and different
\emph{Re} and \emph{p}) to a single empirical function of the form
$\langle\widetilde{P}\rangle=1-exp[-\alpha\overline{\lambda}N-\beta(\overline{\lambda}N)^2]$,
where $\alpha$=0.0018 $\pm$ 1.9 x 10$^{-4}$ and $\beta$= 6.2 x
10$^{-6}$ $\pm$ 2.5 x 10$^{-7}$ are constants. (However, the
constants and could depend on the initial shape of the reaction
interface, a variable we did not explore.)  Note that this equation
(solid line in Fig. 4) fits the data over the entire range
reasonably well. For the lower values of
\emph{$\overline{\lambda}N$}, the quadratic dependence is weak and
we recover Eq.(1). The more rapid approach to saturation at late
times might be due to the transport of initially segregated
reactants to regions where they can react.  A similar transport
effect was demonstrated by Voth et al.~\cite{Voth03} for passive
mixing. Note that we cannot reach the fully reacted state for
$\emph{$\overline{\lambda}$}<3$.
\begin{figure}
\includegraphics[scale=0.45]{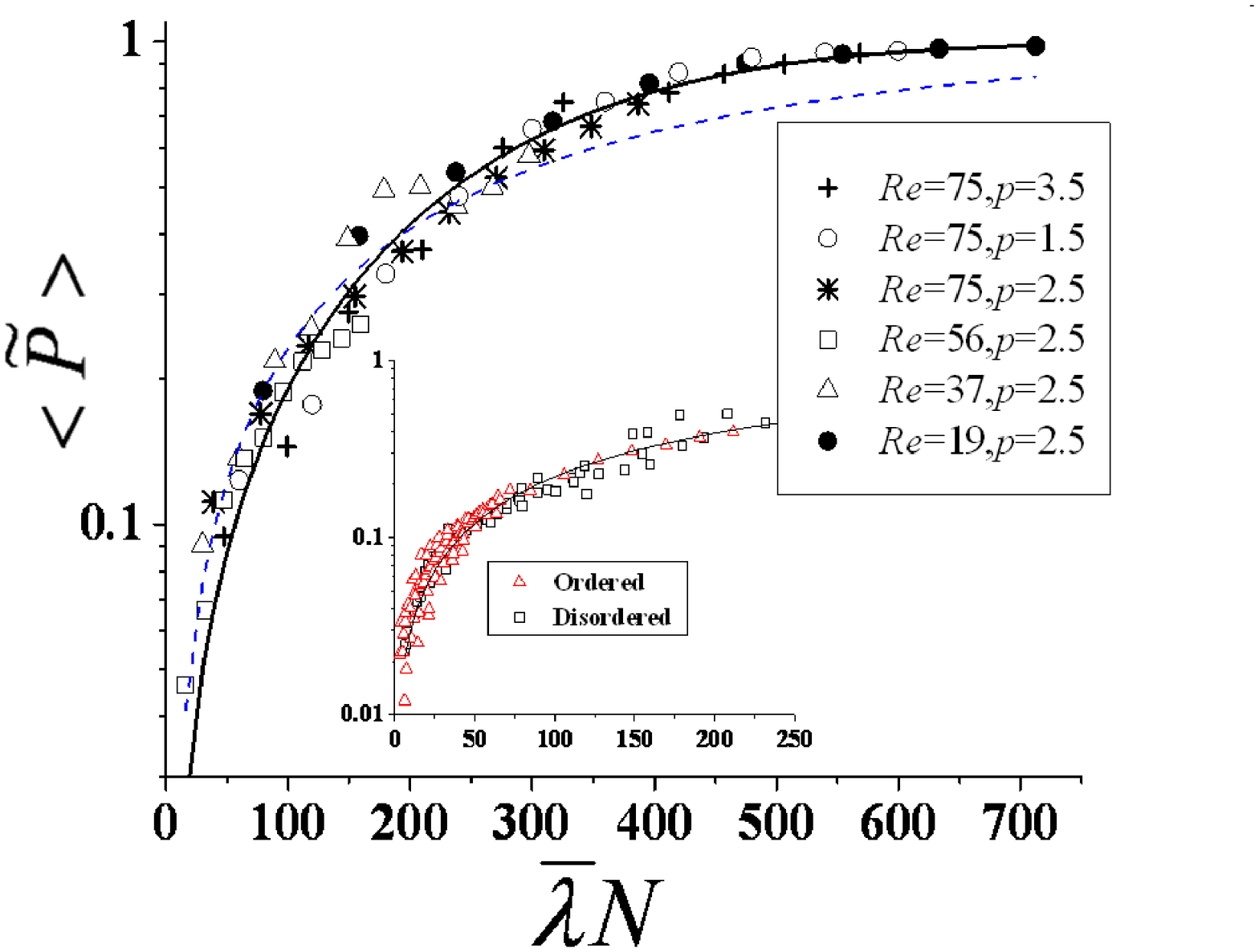}
\caption{\label{fig:epsart} (Color online) Dependence of
$\langle\widetilde{P}\rangle$ on the product of the mean Lyapunov
exponent (\emph{$\overline{\lambda}$}) and period \emph{N} for
disordered flows at various \emph{Re} and \emph{p}. \emph{Insert}:
Early behavior for both ordered and disordered flows (at various
\emph{Re} and \emph{p}). The dotted and solid lines correspond to
Eq.1 (\emph{a}=0.00263) and the empirical fit described in the text,
respectively.}
\end{figure}

In conclusion, we study the interplay of stretching, diffusion, and
fast reaction experimentally, with direct measurement of the
stretching fields. The strong spatial heterogeneity of stretching,
with a distribution spanning many decades,~\cite{Voth02,Arratia05}
causes the early product growth to deviate strongly from the
exponential behavior expected for uniform stretching. Spatial
symmetry, particle displacement, and the departure from time
reversibility affect the reaction progress (and
\emph{$\overline{\lambda}$}). However, the normalized product
concentration up to the fully reacted state approximately follows a
single master curve with \emph{$\overline{\lambda}N$} as the
independent variable, where \emph{$\overline{\lambda}$} is the
measured mean Lyapunov exponent and \emph{N} is the number of mixing
cycles.  This result allows quantitative prediction of the total
product as a function of time for several chaotically mixing flows
with different dynamical features such as spatial symmetry and the
extent of departures from reversibility.  This scaling is not
expected to apply to turbulent flows, but may be appropriate for a
variety of fast reactions that are stirred by chaotic advection.

\begin{acknowledgments}
We thank Z. Neufeld and G. Haller for fruitful discussions, and T.
T\'{e}l and G. K\'{a}rolyi for communicating their theoretical
results to us.  T. Shinbrot provided helpful comments on the
manuscript. This work was supported by NSF DMR-0405187.
\end{acknowledgments}

\bibliography{reaction_refs}

\end{document}